\documentclass{emulateapj}
\usepackage{apjfonts}
\usepackage{graphicx}

\usepackage{amsmath}

\shorttitle{Direct Numerical Simulations of Cosmic-ray Acceleration}
\shortauthors{T. INOUE et al.}

\begin{document}

\title{
Bell Instability-Mediated Diffusive Shock Acceleration at Supernova Blast Wave Shock Propagating in the ISM}
\author{Tsuyoshi Inoue\altaffilmark{1}, Alexandre Marcowith\altaffilmark{2}, and Gwenael Giacinti\altaffilmark{3}}
\altaffiltext{1}{Department of Physics, Konan University, Okamoto 8-9-1, Higashinada-ku, Kobe 658-8501, Japan; tsuyoshi.inoue@konan-u.ac.jp}
\altaffiltext{2}{Laboratoire Universe et Particules de Montpellier (LUPM) Universit\'{e} Montpellier, CNRS/IN2P3, CC72, place Eug\`{e}ne Bataillon, 34095, Montpellier Cedex 5, France}
\altaffiltext{3}{Tsung-Dao Lee Institute and School of Physics and Astronomy, Shanghai 200240, P. R. China}

\begin{abstract}
Supernova blast wave shock is a very important site of cosmic-ray acceleration. However, the detailed physical process of acceleration, in particular, non-linear interplay between cosmic-ray streaming and magnetic field amplification has not been studied under a realistic environment. In this paper, using a unique and novel numerical method, we study cosmic-ray acceleration at supernova blast wave shock propagating in the interstellar medium with well-resolved magnetic field amplification by non-resonant hybrid instability (or Bell instability).
We find that the magnetic field is mildly amplified under typical ISM conditions that leads to maximum cosmic-ray energy $\simeq 30$ TeV for supernova remnants with age $\simeq 1000$ years consistent with gamma-ray observations.
The strength of the amplified magnetic field does not reach so-called saturation level, because cosmic-ray electric current towards the shock upstream has finite spatial extent, by which Bell instability cannot experience many e-folding times.
\end{abstract}


\section{Introduction}
Supernova blast wave shock is known to be an active site of particle acceleration \citep{K95, A08}.
It is widely accepted that the diffusive shock acceleration (DSA) mechanism accounts for the acceleration of charged particles with power-law momentum distribution \citep{BE87, D83}.
Since the rate of galactic supernovae can provide sufficient energy density for cosmic-rays (CRs) below the so-called knee energy ($\simeq3$ PeV), it has been believed that historical young supernova remnants (SNRs) with age $\lesssim 1000$ yr can be PeVatrons.
In order for the young SNRs to be PeVatrons, magnetic field must be amplified at least by two orders of magnitude in upstream of the blast wave shock.
Non-resonant hybrid instability (NRHI or Bell instability) is known to be an effective mechanism of the magnetic field amplification \citep{B04, Marcowith16}.
Near the SNR forward shock, a stream of accelerated CR nuclei (mostly protons) makes an electric current toward upstream that induces a return current of thermal electrons to keep charge neutrality.
The induced return current exerts the Lorentz force on background plasma that amplifies magnetic field by exciting circularly polarized Alfv\'en-like waves.

According to particle-in-cell (PIC) simulations by \cite{CS14b}, the upstream magnetic field can be amplified to $\sim M_{\rm A}^{1/2}$ times the initial strength by the NRHI, where $M_{\rm A}$ is the Alfv\'enic Mach number of the shock.
However, due to a limited scale range of the first-principle simulations, the maximum $M_{\rm A}$ studied by the PIC simulations is only $M_{\rm A}=100$ that leads only an order of magnitude amplification, and we need a case study of realistic $M_{\rm A}\gtrsim 1000$ shock wave.

In order to examine realistic growth of the NRHI, we need to know the spatial structure of the upstream CR current, which is quite hard to compute accurately because the CR current is composed of escaping CR flux whose typical energy is time-dependent.
Hence many authors (sometimes implicitly) assumed the structure of the CR current and discussed the expected level of the upstream magnetic field \citep{SB13, B13, ZPV08, ZP08}.

Observations of young SNRs with ages $\sim 1,000$ yr via synchrotron emission show that the magnetic field downstream of the shock can be amplified to the level of $\sim 100\,\mu$G or more that is clearly stronger than that expected from a shock compression of the interstellar magnetic field, but not enough to make these SNRs PeVatron \citep{VL03, B06, PMBG06}.
However, there is still a possibility for SNRs being PeVatron.
It has been discussed that the maximum CR energy can exceed 1 PeV when the blast wave shock propagates in a dense circumstellar medium (CSM) created by a stellar wind of a progenitor massive star \citep{SB13, MDRTG18}, although the upstream magnetic field amplification by the NRHI is still necessary.

Recently, \cite{I19} developed a novel numerical method that enables self-consistent treatment of the CR current evolution and resulting magnetic field amplification by the NRHI.
Using such a code, \cite{IMGMS21} demonstrated that the blast wave shock propagating in a dense CSM can indeed accelerate particles up to the so-called knee energy $\sim 3$ PeV, if the progenitor mass loss rate before the supernova explosion is high enough.
In their study, it is reported that the NRHI can amplify the magnetic field an order of magnitude from an initial strength, but the final strength immediately before the shock is below the so-called saturation level expected from previous theoretical studies.
They discussed that the moderate amplification by the NRHI is accounted for by a small e-folding number ($\sim 2$--3) due to a limited spatial extent of the CR current.

If the moderate amplification of the NRHI is a common feature of the SNR blast wave shock even in the later ISM propagating phase, it could explain the observed magnetic field strengths of the young SNRs ($\sim 100\,\mu$G in downstream) and explain why they are not PeVatrons.
Thus, in this paper, we examine the NRHI-mediated CR acceleration at the SNR blast wave shock propagating in the ISM.

The paper is organized as follows: in Section 2, we provide the basic equations and numerical settings for simulations. The results of the simulations and their physical interpretation are shown in Section 3. In Section 4, we summarize the paper and discuss the implications.

\section{Basic Equations and Numerical Setups}
\subsection{Basic Equations}\label{BE}
Similar to \cite{IMGMS21}, we solve a hybrid system of the Bell MHD equations and a telegrapher-type diffusion convection equations (or Vlasov–Fokker–Planck equation for the isotropy and first-order anisotropy of CR distribution function) in the polar coordinate around $\theta\sim \pi/2$ \citep[see also, ][]{B13, I19}.
The system equations can be break into the magnetohydrodynamics (MHD) part:
\begin{eqnarray}
&& \frac{\partial\,\rho}{\partial t}+\frac{1}{r^2}\frac{\partial}{\partial r}r^2(\rho\,v_r)=0,\label{eq1}\\
&& \frac{\partial}{\partial t}(\rho\,v_r)+\frac{1}{r^2}\frac{\partial}{\partial r}r^2(\rho\,v_r^{2}+p+\frac{B_\theta^2+B_\phi^2}{8\,\pi})=\nonumber\\
&&\qquad\qquad\qquad\qquad \frac{1}{r}\left\{ \rho(v_{\theta}^2+v_{\phi}^2)+2p+\frac{B_{r}^2}{4\pi}\right\},\label{eq2}\\
&& \frac{\partial}{\partial t}(\rho\,v_\theta)+\frac{1}{r^2}\frac{\partial}{\partial r}r^2(\rho\,v_r\,v_\theta-\frac{B_r\,B_\theta}{4\pi})=\nonumber\\
&&\qquad\qquad\qquad\qquad -\frac{1}{c}j^{({\rm ret})}_{r}\,B_\phi + \frac{1}{r}\left(\frac{B_{r}B_{\theta}}{4\pi}-\rho\,v_{r}\,v_{\theta}\right),\label{eq3}\\
&& \frac{\partial}{\partial t}(\rho\,v_\phi)+\frac{1}{r^2}\frac{\partial}{\partial r}r^2(\rho\,v_r\,v_\phi-\frac{B_r\,B_\phi}{4\pi})=\nonumber\\
&&\qquad\qquad\qquad\qquad \frac{1}{c}j^{({\rm ret})}_{r}\,B_\theta + \frac{1}{r}\left(\frac{B_{r}B_{\phi}}{4\pi}-\rho\,v_{r}\,v_{\phi}\right),\label{eq4}\\
&& \frac{\partial\,\epsilon}{\partial t}+\frac{1}{r^2}\frac{\partial}{\partial r}r^2\{v_r\,(\epsilon+p+\frac{B^2}{8\,\pi}) -B_r\frac{ \vec{B}\cdot\vec{v} }{4\,\pi}\}=-\frac{j^{({\rm ret})}_{r}}{c}(\vec{v}\times\vec{B})_{r}\,, \label{eq5} \\
&& \epsilon=\frac{p}{\gamma-1}+\frac{1}{2}\rho\,v^2+\frac{B^2}{8\pi},\label{eq6}\\
&& \frac{\partial\,B_\theta}{\partial t}=\frac{1}{r^2}\frac{\partial}{\partial r}r^2(B_r\,v_\theta-B_\theta\,v_r)
+ \frac{1}{r}\left(v_{r}B_{\theta}-v_{\theta}\,B_{r}\right),
\label{eq7}\\
&& \frac{\partial\,B_\phi}{\partial t}=\frac{1}{r^2}\frac{\partial}{\partial r}r^2(B_r\,v_\phi-B_\phi\,v_r)
+ \frac{1}{r}\left(v_{r}B_{\phi}-v_{\phi}\,B_{r}\right),
\label{eq8}
\end{eqnarray}
and CR kinetic part:
\begin{eqnarray}
&& \frac{\partial F_0(r,p)}{\partial t}+\frac{1}{r^2}\frac{\partial}{\partial r}\{r^2\,v_r\,F_0(r,p)\}-\frac{1}{3}\frac{\partial\,v_r}{\partial r}\frac{\partial\,F_0(r,p)}{\partial \ln p}\nonumber\\
&&\qquad =-\frac{c}{3}\frac{1}{r^2}\frac{\partial}{\partial r}\{r^2\,F_1(r,p)\}+Q_{\rm inj}(\eta,r,p)\,p^{3},\label{DC1}\\
&& \frac{\partial F_1(r,p)}{\partial t}+\frac{1}{r^2}\frac{\partial}{\partial r}\{r^2\,v_r\,F_1(r,p)\} \\
&&\qquad =-\frac{c}{r}\frac{\partial}{\partial r}\{r\,F_0(r,p)\}
+\frac{c\,F_0}{r}
-\frac{c^2}{3\,\kappa(p,{\vec B})}F_1(r,p),\nonumber \label{DC2}
\end{eqnarray}
where $j^{({\rm ret})}_r$ is the return current density induced by the cosmic ray streaming current ($j^{({\rm ret})}_r=-j^{(\rm cr)}_r$), $Q_{\rm inj}$ is an injection rate, and $\kappa(\vec{B})$ is the diffusion coefficient.
$F_0\equiv f_0\,p^3$ and $F_1\equiv f_1\,p^3$ are, respectively, the isotropic and the first-order anisotropic (drift anisotropy) components of the CR distribution function $f(r,\vec{p})=f_0(r,p)+(p_r/p)\,f_1(r,p)$.
Although they are small corrections, we adequately consider source terms $\propto 1/r$, which are neglected in \cite{IMGMS21}.

The detailed form of $j^{(\rm cr)}_r$, $Q_{\rm inj}$, and $\kappa(\vec{B})$ are given by eqs.~(6)-(9) of \cite{IMGMS21}.
Here, we briefly state how they are described:
The CR current $j^{(\rm cr)}_r$ is computed by momentum space integration of $f_1(p)$.
The injection rate $Q_{\rm inj}$ has a non-zero value only at the shock front, and the rate of injection is determined so that the fraction $\eta$ of the thermal particles go into the DSA process \citep{BGV}.
We employ the following diffusion coefficient \citep{S75, CS14c}
\begin{equation}\label{kappa}
\kappa(p,{\vec B})=\frac{4}{3\,\pi}\frac{\max(B_r^2,\delta B^2)}{\delta B^2}\frac{v_{\rm CR}\,p\,c}{e\,\max(|B_r|,\delta B)},
\end{equation}
where $\delta B^2=B_\theta^2+B_\phi^2$, and $v_{\rm CR}$ is the cosmic-ray velocity at momentum $p$.
Under small amplitude magnetic turbulence ($\delta B \leq B_{r}$), this gives the diffusion coefficient due to pitch angle scattering, while it becomes the Bohm limit coefficient under the amplified field strength for $\delta B>B_{r}$.
This type of diffusion coefficient is supported by the results of PIC simulations \citep{CS14c} and test particle transport calculation in a super-Alfv\'enic turbulence \citep{RII16}.
Note that, since the scale of magnetic field fluctuations induced by the NRHI is roughly 1-2 orders of magnitude smaller than the gyro-scale of the maximum energy CRs, the above diffusion coefficient could be underestimated.
However, it is pointed out that a filamentation instability that grows simultaneously with the NRHI helps to make the fluctuation scale larger \citep{RB12} suggesting the use of eq.~(\ref{kappa}) is reasonable.

Following the method implemented by \cite{IMGMS21}, we numerically treat the momentum space from $p=100$ GeV c$^{-1}$, below which we assume the standard DSA spectrum $f_{0}(r_{\rm sh})\propto p^{-4}$ from the injection momentum.
Thus, our method cannot describe the effect of shock structure modification by cosmic-ray pressure, which is carried by CRs with $p\sim 1$ GeV.
In this paper, we study the cases with the injection rate $\eta =0.1-1.0\times 10^{-4}$ in which the effect of CR pressure is not substantial.
According to a theoretical modeling of multi-wavelength emission of a historical young SNR SN1006 by \cite{BKV12}, $\eta \simeq 10^{-4}$ gives the best fit.
Note that studying the effect of the CR pressure under a larger injection rate is still important, because the injection rate can be locally enhanced, and we will further study on it in our future papers.

\subsection{Fluid Initial Conditions}\label{inicnd}
The inner and outer boundaries of the numerical domain are located at $r_{\rm in}=0.5$ pc and $r_{\rm out}=3\times 10^{19}$ cm $\simeq 10$ pc for most models.
To induce a blast wave shock, we initially set an ejecta of radius $R_0=1.0$ pc, velocity $v_0=10^4 (r/R_0)$ km s$^{-1}$, and density $\rho_0=2.1\times 10^{-23}$ g cm$^{-3}$.
This leads the total kinetic energy of the ejecta to be $E_{\rm kin}=0.76\times 10^{51}$ erg.

The initial ISM is distributed in $R_0<r<r_{\rm out}$ with uniform density $\rho_{\rm ISM}=m\,n_{\rm ISM}$ and temperature $T_{\rm ISM}=10^6$ K, where $m=1.27\,m_{\rm p}$ is the mean gas particle mass.
To study the impact of the ISM density on CR acceleration, we examine three types of the ISM density $n_{\rm ISM}=0.5,\,0.1,$ and 0.02 cm$^{-3}$.
For the lowest ISM density model ($n_{\rm ISM}=0.02$ cm$^{-3}$), we use $r_{\rm out}=4\times 10^{19}$ cm because the shock propagates more distant than the larger density models.
Given that the ISM temperature $T_{\rm ISM}=10^6$ K, the ISM sound speed is calculated to be $c_{\rm s}\simeq 100$ km s$^{-1}$.
Thus, the Mach number of the forward shock becomes $\sim v_0/c_{\rm s} \simeq 100$ for the early free expansion stage.
Note that our choice of the ISM temperature is realistic for the lower ISM density models of $n_{\rm ISM}=0.1$ and 0.02 cm$^{-3}$, but for $n_{\rm ISM}=0.5$ cm$^{-3}$ case, the realistic temperature would be around $10^4$ K that leads to a sonic Mach number around 1000.
In general, it is numerically hard to stably follow the dynamics of the shock with Mach number 1000\footnote{We employ HLLD Riemann solver \citep{MK05} in the integration of the MHD part, which is known to be a very robust scheme.}.
Thus, we choice $T_{\rm ISM}=10^{6}$ K even for $n_{\rm ISM}=0.5$ cm$^{-3}$ model.
Since the upstream temperature does not influence the NRHI dynamics, the result seems to be unchanged even if we set a lower ISM temperature.

We assume the constant radial component of the magnetic field $B_r=3\,\mu$G for most models.
\footnote{Since we set inconsistent radial dependence of $B_r$ to the divergence-free condition of the magnetic field, which brings erroneous force term $\propto 1/r$ to the Lorentz-force.
However, this erroneous term is orders of magnitude smaller than other physical terms whose ratio is estimated to be $\lesssim \lambda_{\rm NRH}/r_{\rm in}\sim 10^{-5}$.
Thus, our initial magnetic field condition does not raise any noticeable issue.}.
The upstream Alfv\'en velocity is calculated to be 
$v_{\rm A} = B_{r}/\sqrt{4\pi\rho_{\rm ISM}}=8.2\mbox{ km s}^{-1}(B_r/3\mu\mbox{G})(n_{\rm ISM}/0.5\mbox{cm}^{-3})^{-1/2}$.
For $B_{\theta}$ and $B_{\phi}$, we set turbulent fluctuations by superposing Alfv\'en waves with flat power spectrum ($B_{k}^2=$const.).
The fluctuation dispersion is set as $\xi_B\equiv \langle (B_{\theta}^2+B_{\phi}^2)/B_r^2 \rangle=0.1$ for most runs, and we also examine $0.01$ case to study the effect of seed amplitude on the NRHI growth.
We show a summary of our numerical model parameters in Table 1.


\subsection{Boundary Conditions}
For the boundary conditions of the MHD part of equations, we use the free boundary conditions, except we set $\vec{v}=0$, and $B_{\theta}=B_{\phi}=0$ at the inner boundary.
This set of boundary conditions reproduces the blast wave evolution starting from the free expansion phase ($v_{\rm sh}\propto t^0$) and then shift to the Sedov-Taylor phase ($v_{\rm sh}\propto t^{-3/5}$; see Fig.~\ref{f3a1} below).

For the spatial boundaries of the CR part, we assume that CRs do not penetrate into the ejecta by setting $\kappa=0$ in the ejecta.
At the outer spatial boundary, outgoing free boundary conditions are imposed: $F_0(p,r_{\rm out}+\Delta r)=F_0(p,r_{\rm out})$ and $F_1(p,r_{\rm out}+\Delta r)={\rm max}[\,F_1(p,r_{\rm out}),\,0\,]$.
For the momentum space, as stated in \S 2.1, we assume $f_0\propto p^{-4}$ below the boundary at $p=100$ GeV c$^{-1}$ (see, eq.~(15) in \cite{IMGMS21} for implementation), while for $f_1$, we impose null values outside the numerical domain of [100 GeV c$^{-1}$,10 PeV c$^{-1}$].

\begin{table*} \label{t1}
\caption{Model Parameters}
\begin{center}
\scalebox{1.}[1.]{
\begin{tabular}{c|ccccc|cc}
\hline\
Model ID & $n_{\rm ISM}$ & $B_{r}$ & $\xi_{B,\rm ini}$ & $\eta$ & Bell term$^{a}$ & $E_{\rm cut}$ & $\epsilon_{B}$$^{b}$\\  \hline
1 & $0.5$ & 3.0$\mu$G & 0.1 & $1\times 10^{-4}$ & yes & $31\pm 1.2$ TeV & $4.0\times 10^{-2}$ \\
2 & $0.1$ & 3.0$\mu$G & 0.1 & $1\times 10^{-4}$ & yes & $41\pm0.7$ TeV & $3.6\times 10^{-2}$ \\
3 & $0.02$ & 3.0$\mu$G & 0.1 & $1\times 10^{-4}$ & yes & $29\pm0.6$ TeV & $1.3\times 10^{-2}$ \\
4 & $0.5$ & 3.0$\mu$G & 0.01 & $1\times 10^{-4}$ & yes & $29\pm0.9$ TeV & $5.0\times 10^{-2}$ \\
5 & $0.5$ & 10.0$\mu$G & 0.1 & $1\times 10^{-4}$ & yes & $37\pm1.3$ TeV & $12.6\times 10^{-2}$ \\
6 & $0.5$ & 3.0$\mu$G & 0.1 & $3\times 10^{-5}$ & yes & $19\pm0.5$ TeV & $1.5\times 10^{-2}$ \\
7 & $0.5$ & 3.0$\mu$G & 0.1 & $1\times 10^{-5}$ & yes & $9.3\pm0.4$ TeV & $0.24\times 10^{-2}$ \\
8 & $0.5$ & 3.0$\mu$G & 0.1 & $1\times 10^{-4}$ & no & $0.82\pm0.02$ TeV & $2.4\times 10^{-5}$ \\
\hline\
\end{tabular}
\tablenotetext{a}{if no, we always set null return current $j_{r}^{({\rm ret})}=0$.}
\tablenotetext{b}{Ratio of downstream magnetic energy density to upstream kinetic energy. The downstream magnetic energy is averaged over 1000 $\Delta x$ region from the shock front.}
}
\end{center}
\end{table*}

\subsection{Numerical Resolution}\label{RES}
Since the NRHI growth capturing is the major purpose of this study, we need to resolve the most unstable scale of the instability \citet{B04}: $\lambda_{\rm B}=c\,B_{r}/|j_r^{({\rm CR})}|$.
The critical scale below which the NRHI is stabilized by magnetic tension force is given by $\lambda_{\rm B}/2$.
The results of our fiducial model (Model 1) show that, although it depends highly on time and distance from the shock front, the CR current density roughly takes a value of $j^{(\rm{cr})}_r\sim 10^{-10}$ esu s$^{-1}$ cm$^{-2}$ near the shock front.
Using this current density, the most unstable spatial and timescales of the NRH instability can be estimated as
\begin{eqnarray}
\lambda_{\rm B}&\simeq& 0.9\times 10^{15} \mbox{ cm} \,\left( \frac{j^{({\rm cr})}}{10^{-10}\,\mbox{esu s$^{-1}$cm$^{-2}$}} \right)^{-1}\left( \frac{B_r}{3\,\mu\mbox{G}} \right), \label{kB}\\
\omega_{\rm B}^{-1}&=&\frac{\lambda_{\rm B}}{2\pi\,\,v_{\rm A}}=\frac{c\,\rho^{1/2}}{\pi^{1/2}\,|j^{(\rm{cr})}_r|} \nonumber\\
&\simeq& 5.5 \mbox{ yr} \,\left( \frac{j^{({\rm cr})}}{10^{-10}\,\mbox{esu s$^{-1}$cm$^{-2}$}} \right)^{-1}\left( \frac{n}{0.5\,\mbox{cm}^{-3}} \right)^{1/2}. \label{tB}
\end{eqnarray}
According to the results of test simulations by \cite{IMGMS21}, the numerical dispersion relation can reproduce the analytic one if $\lambda_{\rm B}$ is resolved by more than 32 numerical cells.
To satisfy the above condition, we resolve whole spatial numerical domain by $N_{\rm cell}=2^{20}\simeq 10^6$ cells, leading to the spatial resolution of $\Delta r = (r_{\rm out}-r_{\rm in})/N_{\rm cell}\simeq 3\times 10^{13}$ cm.
Thus, the ratio of the instability scale and the resolution is estimated to be $\lambda_{\rm B}/\Delta r\simeq 30$ that satisfies the resolution requirement by \cite{IMGMS21}.

For the momentum space, we explicitly treat a range of [$p_{\rm L}=100$ GeV c$^{-1}$, $p_{\rm U}=1$ PeV c$^{-1}$], which is divided into uniform $N_{{\rm cell},p}=64$ numerical cells in the logarithmic scale, i.e., $\Delta \ln p=\ln (p_{\rm U}/p_{\rm L})/N_{{\rm cell},p}$.
Because most CR current is composed of CRs with $E\gtrsim 10$ TeV (see Fig.~\ref{f3a3} below), as long as the lower boundary of the momentum space set below 1 TeV/$c$, the result would not be affected by the boundary.
This resolution is tested to be enough for numerical convergence \citep[see, \S~2.6 of][]{IMGMS21}.

\subsection{Advantage of Our Numerical Scheme}
The detailed numerical method to solve the basic eqs.~(\ref{eq1})-(\ref{DC2}) is described in \S~2.5 of our previous paper \citep{IMGMS21}.
Here we briefly state the advantages of our method:
If we employ the usual diffusion equation and solve it explicitly, a required time-step for numerical stability becomes $\Delta t\le 0.5\,\Delta r^2/\kappa(B_r,p_{\rm U})\sim 10^{-3}$ sec for the initial state, indicating a thousand-year scale integration is impossible even with the largest scale super-computer.

On the other hand, our hyperbolic type basic equations require $\Delta t\le C_{\rm CFL}\,\Delta r/(c/\sqrt{3})\sim 10^{3}$ sec that is still short but the thousand year integration is executable with a modern super-computer.
Here, $c/\sqrt{3}$ is the free streaming velocity of CRs, and $C_{\rm CFL}$ is the CFL number that is set to be 0.8.

In addition to the alleviated timestep, the hyperbolic equation generally has good compatibility with parallel computing, enhancing the advantage of our method.
One may think of implicit schemes or super-time-stepping methods for integration, which allow large timestep.
However, the implicit scheme is incompatible with parallel computing and has insufficient accuracy to recover the DSA spectrum.
Super-time-stepping can only enlarge $\Delta t$ by roughly an order of magnitude, if we want to keep the appropriate cosmic-ray spectrum.
Thus, it cannot be used under realistic situation considered in this paper.

\section{Results} \label{Sec:Rslt}
\subsection{Result of Model 1 - 3}

\begin{figure}[t]
\includegraphics[scale=0.65]{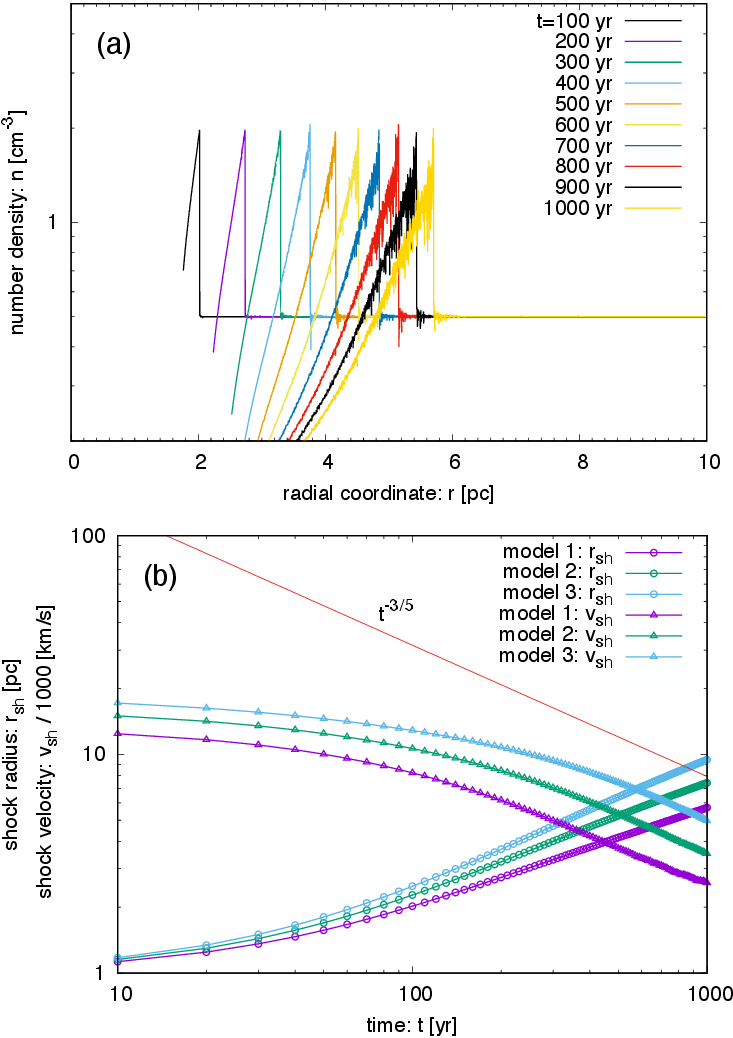}
\caption{\label{f3a1}
Panel (a): density structure for Model 1 (fiducial model).
Different colors show different snapshot times.
Panel (b): forward shock position (circles) and shock velocity divided by 1000 km s$^{-1}$ (triangles) for three different models.
Thin red line is a reference line to the Sedov-Taylor solution: $\propto t^{-3/5}$.
}\end{figure}

\begin{figure*}[t]
\includegraphics[scale=1.]{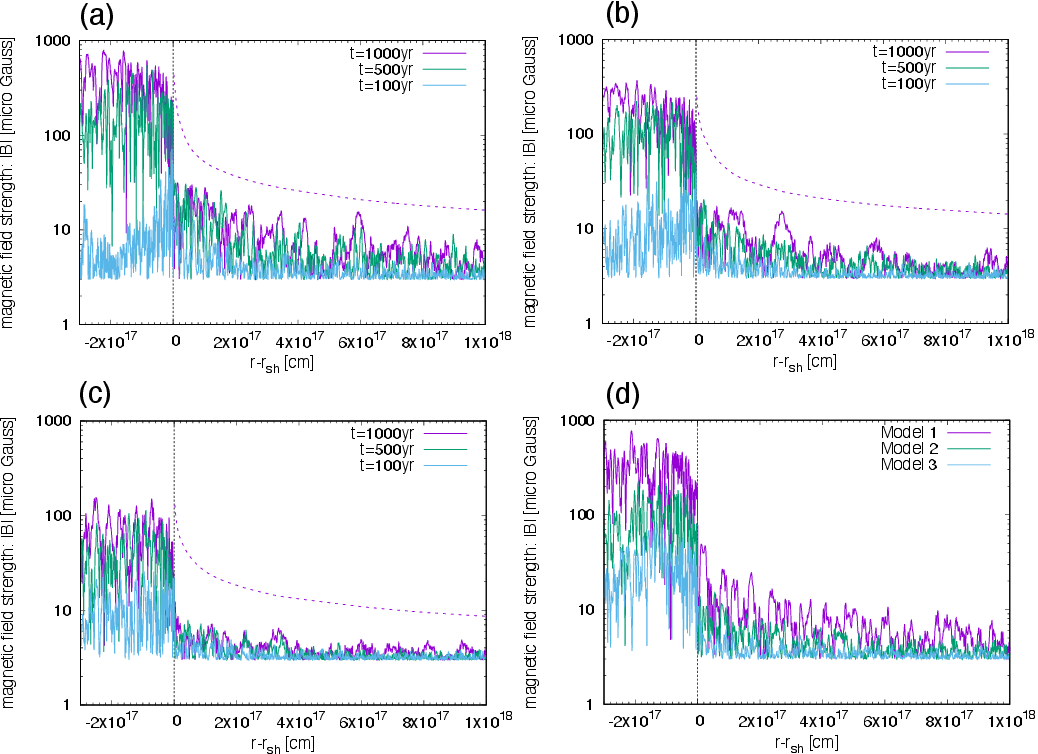}
\caption{\label{f3a2}
Panel (a): structure of the magnetic field strength around the shock at $t=$100, 500, and 1000 yr for Model 1.
Panel (b): Same as panel (a) but for Model 2.
Panel (c): Same as panel (a) but for Model 3.
Panel (d): magnetic field strength of the results of Model 1-3 when the shock passes $r=5$ pc.
Dashed lines in panels (a)-(c) represent the saturation amplitude of NRHI given by eq. (\ref{eq:Bsat}) for $t=1000$ yr.
}\end{figure*}

\begin{figure}[t]
\includegraphics[scale=0.6]{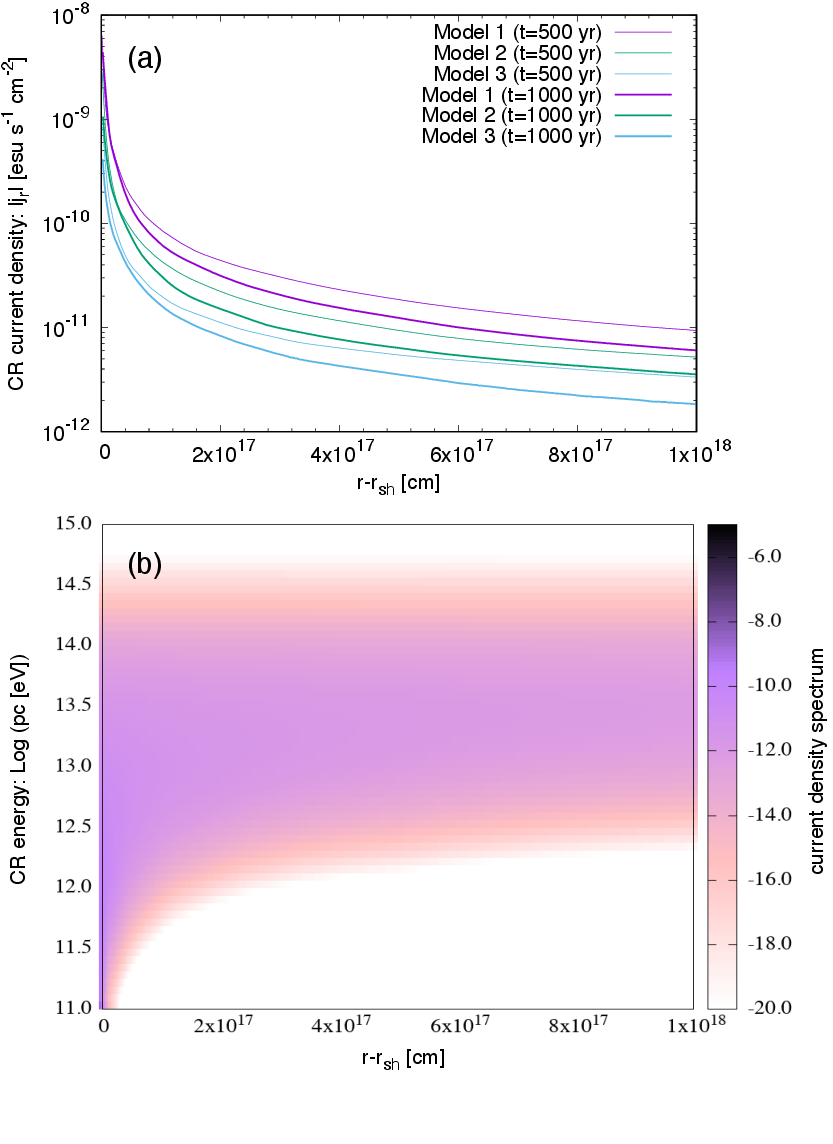}
\caption{\label{f3a3}
Panel (a): structure of the CR current densities for Model 1, 2 and 3 at $t=500$ (thin lines) and 1000 yr (thick lines).
Panel (b): upstream CR current spectrum $j_p(r,p)=dj_r^{({\rm CR})}/d\ln p$ at $t=1000$ yr for Model 1.
}\end{figure}

\begin{figure}[t]
\includegraphics[scale=0.65]{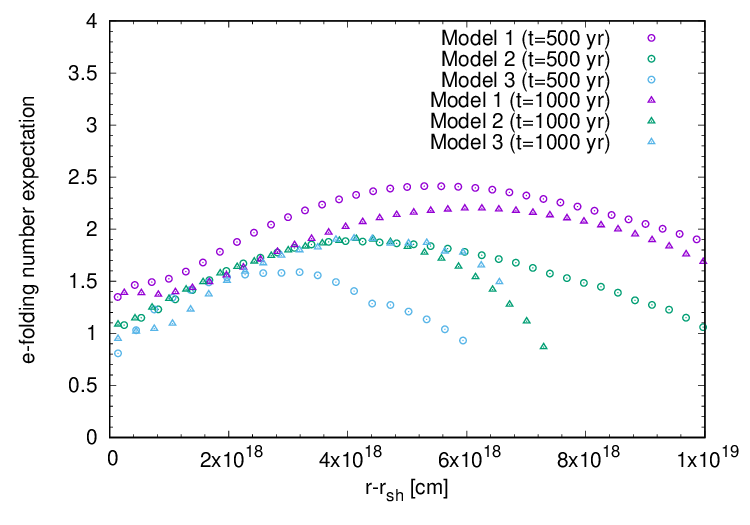}
\caption{\label{f3a4}
Expected e-folding number of the NRHI at distance $r-r_{\rm sh}$ (see eq.~(\ref{e-fold})) for Model 1 (magenta), Model 2 (green), and Model 3 (light blue) at $t_0=500$ yr (circle) and 1000 yr (triangle).
}\end{figure}

\begin{figure}[t]
\includegraphics[scale=0.7]{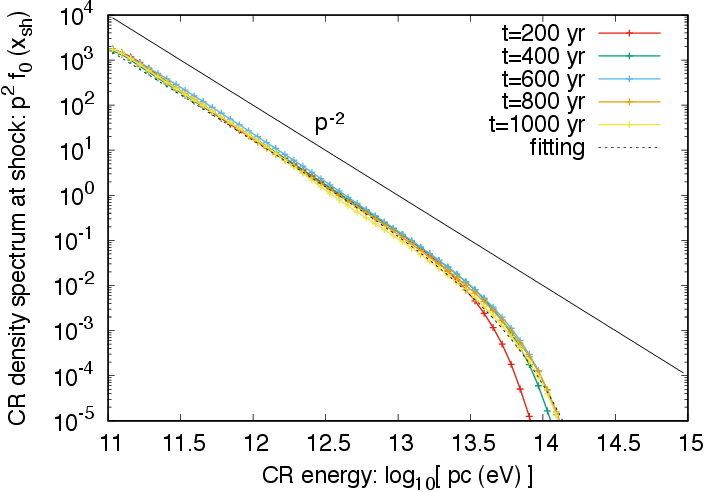}
\caption{\label{f3a5}
CR density spectrum ($n_p=f_0\,p^2$) of Model 1 immediately behind the shock.
Black line represents the DSA spectrum ($\propto p^{-2}$).
We can reasonably fit the spectrum by a power law function with the standard DSA spectral index and exponential cut-off: $n_p\propto p^{-2}\,\exp(-pc/E_{\rm cut})$.
The result of fitting for $t=1000$ yr is shown as a dotted line ($E_{\rm cut}=31$ TeV).
}\end{figure}

Panel (a) of Figure \ref{f3a1} shows the density structures of the result of Model 1 (fiducial model), where we only plot outside of the contact discontinuity.
We see that the density structure around the shock becomes very noisy.
This is due to the back reaction of magnetic field amplification.
Panel (b) shows the forward shock velocity evolution, where the shock position is defined as the outermost position of $v_{r}>1,000$ km s$^{-1}$.
We can confirm that the shock evolves from the initial free expansion phase ($v_{\rm sh}\propto t^0$) to the Sedov-Taylor phase ($v_{\rm sh}\propto t^{-3/5}$).
As expected, the larger the initial ISM density, the faster the shock attenuates.

Figure \ref{f3a2} shows the structure of the magnetic field strength around the shock at $t=$100, 500, and 1000 yr for Model 1 (panel a), Model 2 (panel b), and Model 3 (panel c).
Model 1-3 have similar levels of the magnetic field on the order of tens of micro-Gauss immediately ahead of the shock and a hundred micro-Gauss behind the shock.
To make a more detailed comparison, in panel (d) of Figure \ref{f3a2}, we plot the field strength of the results of Model 1-3 when the shock passes $r=5$ pc.
In addition, in Table 1, we list values of so-called $\epsilon_{B}$, i.e., the ratio of downstream magnetic energy density to upstream kinetic energy, where the downstream magnetic energy is averaged over 1000 $\Delta x$ region from the shock front at $t=1000$ yr.
We see that the degree of the amplification increases with the ISM density.
This is naturally accounted for by a larger total number of CRs, which is proportional to the upstream density, and thus stronger CR current.

To see how magnetic fields grow in the upstream region, we plot the structure of the CR current density in the top panel of Figure \ref{f3a3}.
We also show the upstream CR current spectrum $j_p=dj_r^{({\rm CR})}/d\ln p$ at $t=1000$ yr for Model 1 in the bottom panel.
We see that CRs with energy $pc\sim 10$ TeV mostly compose the upstream CR current.
The spatial extent of the CR current is an order of magnitude larger than the diffusion length of the 10 TeV CRs:
\begin{eqnarray}
l_{\rm diff}&=&\kappa(p)/v_{\rm sh}\nonumber\\
&\simeq& 10^{17}\mbox{ cm}\,\,\xi_B^{-1}\left(\frac{p_{\rm max}}{10\,\mbox{TeV}\,c^{-1}}\right) \left(\frac{B}{10\,\mu\mbox{G}}\right)^{-1} \left(\frac{v_{\rm sh}}{3\times10^3\,\mbox{km s}^{-1}}\right)^{-1},
\end{eqnarray}
At time $t_0$, advection time $t_{\rm ad}$ for the gas at distance $l=r-r_{\rm sh}$ from the shock front is obtained by solving 
\begin{equation}
l=\int_{t_0}^{t_{\rm ad}}v_{\rm sh}\,dt,
\end{equation}
For $t_0\gtrsim 500$ yr, from the Sedov-Taylor solution, the shock velocity is given by $v_{\rm sh}=v_{\rm sh}(t_0)(t/t_0)^{-3/5}$, which leads
\begin{equation}
t_{\rm ad}=\left( \frac{2}{5}\frac{l}{v_{\rm sh}(t_0)\,t_0}+1 \right)^{5/2}t_0.
\end{equation}
Using $t_{\rm ad}$, the expected e-folding number of the NRHI at distance $l$ is computed as
\begin{equation}\label{e-fold}
\sigma(t_0,r) = (t_{\rm ad}-t_0)\,\omega_{\rm B}.
\end{equation}
In Figure \ref{f3a4}, we plot $\sigma$ at $t_0=500$ yr and 1000 yr for Model 1, 2, and 3.
An important overall feature is that, as reported by \cite{IMGMS21}, the NRHI has a small e-folding number (1.5-2.5) that can amplify the seed fluctuations only by roughly factor $e^{\mbox{ a few}}\sim10$.
The expected e-folding number becomes smaller as ISM density decreases (as model number increases from 1 to 3).
This seems to be reasonable behavior because, at fixed time, the total number of injected particles is a decreasing function of the initial ISM density that leads to smaller escape particles for smaller ISM density cases.

It has been expected that, if the e-folding number is large enough, the NRHI grows until magnetic field strength reaches the so-called saturation level, which happens once the gyroradius of maximum energy CRs becomes smaller than the NRHI critical scale \citep{B04}:
\begin{eqnarray}
B_{\rm sat}&=&\left(\frac{4\pi\,j_{r}^{({\rm CR})}\,p_{\rm max}}{e}\right)^{1/2}\nonumber\\
&\simeq& 64\,\mu\mbox{Gauss}\,\left(\frac{j_{r}^{({\rm CR})}}{10^{-10}\,\mbox{esu s}^{-1}\,\mbox{cm}^{-2}}\right)^{1/2}
\,\left(\frac{p_{\rm max}}{30\,\mbox{TeV}\,c^{-1}}\right)^{1/2}, \label{eq:Bsat}
\end{eqnarray}
where $p_{\rm max}$ is the maximum (escaping) CR momentum.
The magnetic field strengths in the simulations do not reach the above saturation level.
This result is very similar to the case of an earlier phase of SNR studied by \cite{IMGMS21}.
In panels (a)-(c) of Figure \ref{f3a2}, we plot the saturation amplitude (eq. [\ref{eq:Bsat}]) estimated using local $j_{r}^{({\rm CR})}$ and the fitted $E_{\rm cut}$ for $t=1000$ yr data as dashed lines.
Previous simple theoretical modelings assumed enough e-folding number for the NRHI to saturate.
However, if we take into account the realistic spatial extent of the cosmic-ray current, it does not reach saturation in the SNR environment.

Figure \ref{f3a5} shows the CR density spectrum ($n_p=f_0\,p^2$) of Model 1 immediately behind the shock front.
We can reasonably fit the spectrum by a power law function with the standard DSA spectral index and exponential cut-off: $n_p\propto p^{-2}\,\exp(-pc/E_{\rm cut})$.
In Table 1, we list the cut-off energies $E_{\rm cut}$ at $t=1,000$ yr for all models as a result of the fitting.
The resulting cut-off energies become $E_{\rm cut}\sim 10$ TeV at $t=1,000$ yr for all models, except no magnetic field amplification model (Model 8).
Since the larger ISM density model (Model 1) has slightly stronger magnetic field strength compared to Model 2 and 3 but has smaller shock velocities, these complementary effects seem to make similar maximum energies between Model 1 to Model 3.
The numerical cut-off energy is roughly consistent with the maximum energy of the age-limited DSA acceleration theory under the amplified magnetic field strength \citep{BE87}:
\begin{equation}
E_{\rm max}\simeq 43\mbox{ TeV}\,\xi_B \left(\frac{B}{10\,\mu\mbox{G}}\right) \left(\frac{v_{\rm sh}}{3000\,\mbox{km s}^{-1}}\right)^2 \left(\frac{t}{1000\,\mbox{yr}}\right).
\end{equation}
We can also understand the maximum energy of Model 8, which artificially suppresses the NRHI, if we substitute unamplified magnetic field strength of $3\,\mu$G with $\xi_B=0.1$.

\begin{figure}[t]
\includegraphics[scale=0.7]{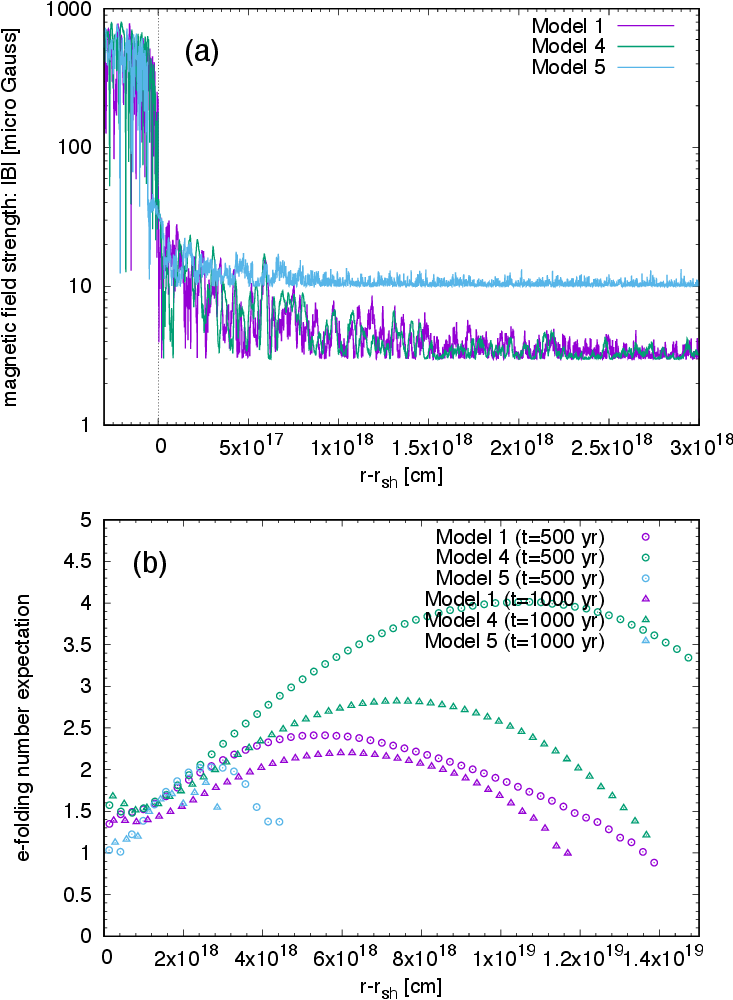}
\caption{\label{f3b1}
Panel (a): magnetic field structure around the shock for Model 1 ($B_r=3\,\mu\mbox{G},\xi_B=0.1$), Model 4 ($B_r=3\,\mu\mbox{G},\xi_B=0.01$) and 5 ($B_r=10\,\mu\mbox{G},\xi_B=0.1$) at $t=1000$ yr.
Panel (b): Expected e-folding number of the NRHI at distance $r-r_{\rm sh}$ (see eq.~(\ref{e-fold})) for Model 1 (magenta), Model 4 (green), and Model 5 (light blue) at $t_0=500$ yr (circle) and 1000 yr (triangle).
}\end{figure}

\subsection{Influence of Seed Amplitude and Initial Field Strength}
In Model 4, we run the case that has a smaller initial magnetic field fluctuation amplitude ($\xi_B=0.01$) than the fiducial model.
We also study the case of stronger initial field strength ($B_r=10\,\mu$G) in Model 5.
In the top panel of Figure \ref{f3b1}, we compare the magnetic field structure around the shock for Model 1, 4, and 5 at $t=1000$ yr.
We see that the final field strength does not substantially depend on the initial seed amplitude and field strength.
This is because the small upstream fluctuations allow more CR flux that results in active NRHI growth in Model 4, while the stronger initial field in Model 5 has a negative effect on the CR stream.
Indeed, from the bottom panel of Figure \ref{f3b1} that is the same as Figure \ref{f3a4} but for Model 1, 4, and 5, we can confirm that the results of smaller initial seed model show larger e-folding number, and larger initial strength model shows more compact NRHI active region.

The results of our self-consistent calculation of CR current structure indicate that the e-folding number depends highly on the degree of upstream magnetic field fluctuations, but the final magnetic field strength in the immediate shock upstream takes a similar value in a realistic range of parameters explored in this paper.
The insensitivity of the magnetic field strength around the shock on the upstream seed fluctuations seems reasonable, if we notice that total electric charge escapes to the upstream is the most crucial quantity, which is constant between Model 1, 4, and 5 owing to the common injection rate and upstream density.
As a consequence of the similar magnetic field strengths, the cut-off CR energies take similar values in Model 1, 4, and 5 (see, Table 1).

\begin{figure}[t]
\includegraphics[scale=0.7]{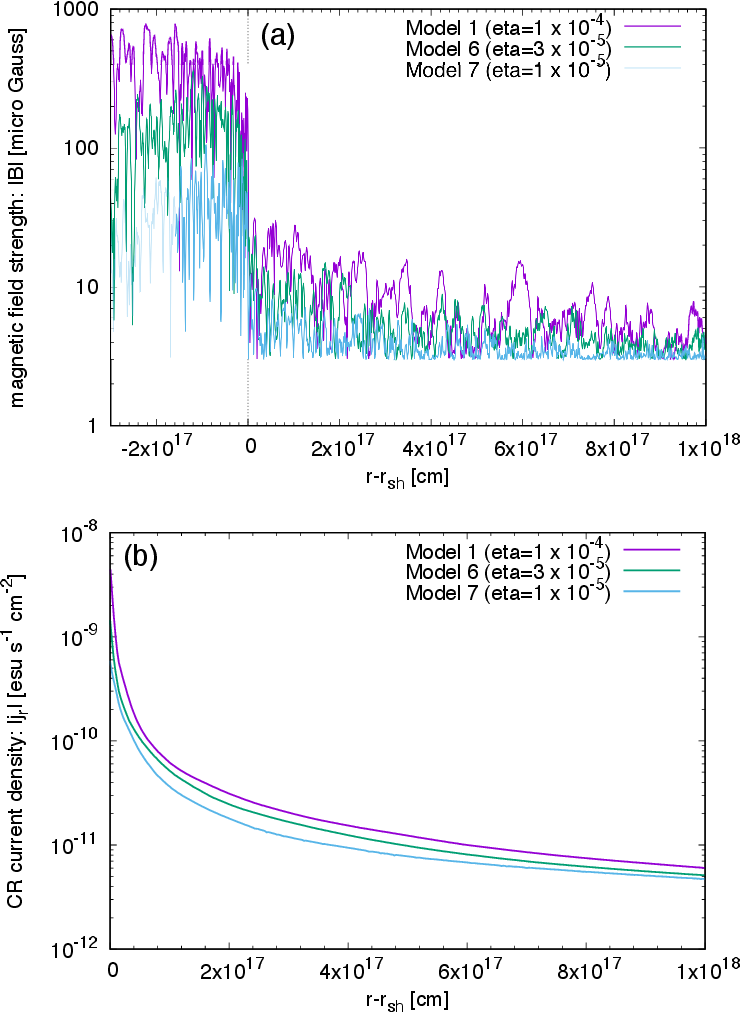}
\caption{\label{f3c1}
Panel (a): magnetic field structure around the shock for Model 1 ($\eta=1\times 10^{-4}$), Model 6 ($\eta=3\times 10^{-5}$) and Model 7 ($\eta=1\times 10^{-5}$) at $t=1000$ yr.
Panel (b): structure of the CR current densities for Model 1, 6, and 7 at $t=1000$ yr.
}\end{figure}

\begin{figure}[t]
\includegraphics[scale=0.7]{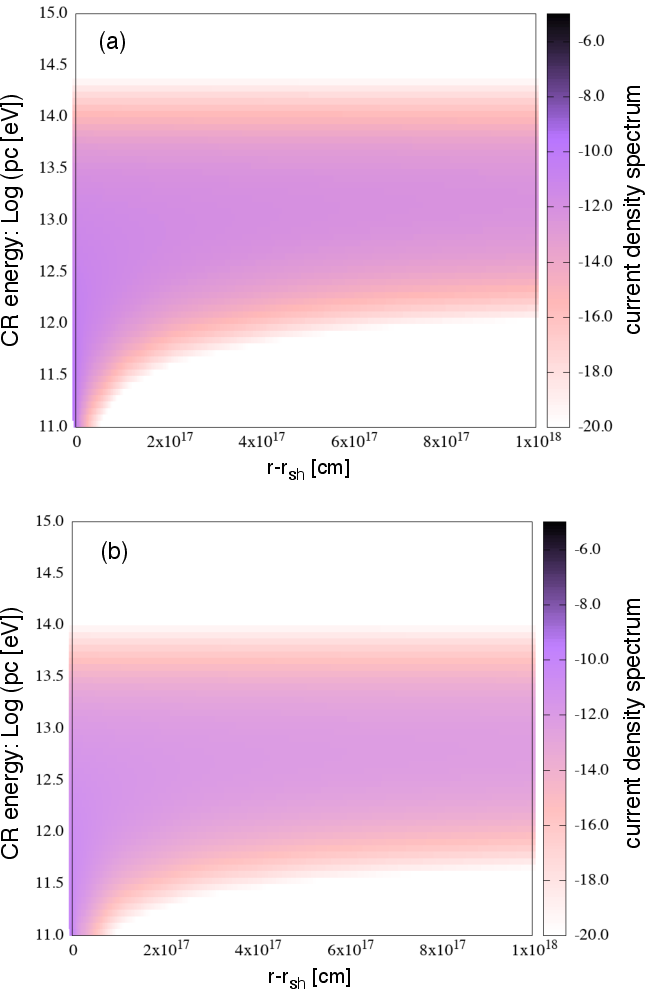}
\caption{\label{f3c2}
Panel (a): upstream CR current spectrum $j_p(r,p)=dj_r^{({\rm CR})}/d\ln p$ at $t=1000$ yr for Model 6.
Panel (b): same as panel (a) but for Model 7.
}\end{figure}

\subsection{Effect of the Injection Rate}
Since the CR current has a substantial impact on the growth rate of NRHI, the injection rate $\eta$ could have a strong influence on the magnetic field amplification and the resulting maximum CR energy.
Figure \ref{f3c1} shows the structures of magnetic field strength (top) and CR current (bottom) around the shock for Model 1, 6, and 7 at $t=1000$ yr.
We see that, although the amplification is more effective as $\eta$ increases, the influence of $\eta$ variation is not so striking, even though there is an order of magnitude difference of $\eta$ between Model 1 and Model 7.
This somewhat similar magnetic field amplification level results in similar cut-off energies (only factor three difference) between Model 1, 6, and 7 (see Table 1).

This small dependence on $\eta$ can be simply explained as follows:
In the very early stage, the larger $\eta$ model produces a stronger CR current that induces active NRHI and boosts up $E_{\rm max}$ faster than the smaller $\eta$ model.
However, since CR current is composed of CRs around $E_{\rm max}$, it is weakened as $E_{\rm max}$ enlarges.
Because CR density at shock obeys $dn_p\propto p^{-2}\,dp$, factor two enhancement of cut-off energy leads to a quarter CR density around cut-off energy and thus reduced escaping CR current.
As a consequence, in a later stage, the strength of the CR current becomes similar level in all Model 1, 6, and 7 resulting in not substantially different magnetic field strengths.
Figure \ref{f3c2} shows CR current spectra for Model 6 (top) and Model 7 (bottom) from which we see that CR current in Model 6 is composed of higher energy CRs than that of Model 7.

\section{Summary and Discussion}
We have studied CR acceleration at a supernova blast wave shock propagating in the ISM under the influence of NRHI.
Our method handles non-linear interplay between CR acceleration, CR streaming, and magnetic field amplification with sufficient resolution.
Our findings can be summarized as follows:
\begin{itemize}
    \item The NRHI is indeed effective at the young SNR forward shocks with age around 1,000 yr irrespective of the ISM density.
    Downstream $\epsilon_B$ (the density ratio of upstream kinetic energy and downstream magnetic energy) can be 1-10\%, while it becomes 0.002\% if the effect of NRHI is switched off.
    \item The level of the NRHI amplification does not reach the so-called saturation level, at which the gyroradius of escaping CRs becomes smaller than the critical length of the NRHI. This is because the realistic spatial extent of the upstream CR current is finite, while previous theories assumed sufficiently large extent to achieve the saturation. 
    \item The amplified strength of the magnetic field at $t=1,000$ yr does not substantially depend on the ISM density, because smaller ISM density leads to faster shock propagation that compensates the CR current strength.
    \item The magnetic field strength at immediate shock upstream does not substantially depend on the initial field strength and the seed fluctuation amplitude. This is because smaller initial strength and/or seed fluctuations lead to more active amplification via more spatially extended CR current (due to larger CR diffusion coefficient).
    \item Although larger injection rate models show more effective amplification and larger resulting maximum energy of CRs, they show weaker dependence to the injection rate than the linear relation (an order of magnitude change of the injection rate results in only a factor around three difference in the maximum energy). This seems to stem from the fact that the escaping CR number (current strength) is a decreasing function of the maximum energy.
    \item The above mentioned somewhat stable nature of the NRHI amplification against the environmental conditions always leads similar maximum CR energy of $\sim 30$ TeV at $t=1,000$ yr. This is a somewhat unexpected result, but consistent with recent observations of young SNRs with $t_{\rm age }\lesssim 1,000$ yr \citep{SBYO22}.
\end{itemize}

Very recently, Tibet AS$\gamma$ experiment and Large High Altitude Air Shower Observatory (LHAASO) found PeV/sub-PeV gamma-rays potentially from some young and middle-aged SNRs ($t_{\rm age}\sim 10^3-10^4$ yr) \citep{ASG21, LHASSO21}.
Given the results of the present study, these ultra-high energy gamma-rays seem to be not from SNR shells, but from escaped CRs that are accelerated much earlier phase of the SNR or from other nearby high-energy objects.
However, since we have only studied the cases of moderate CR injection rate ($\eta\sim 10^{-4}$ at maximum), there may be an SNR population that has a higher injection rate by yet unknown effect.
In such a higher injection rate model, non-negligible CR pressure would lead shallower CR spectrum and thus more efficient growth of the NRHI, which will be studied in our future works.

\acknowledgments
The numerical computations were carried out using the supercomputer "Flow" at Information Technology Center, Nagoya University and XC50 system at the Center for Computational Astrophysics (CfCA) of National Astronomical Observatory of Japan.
This work is supported by Grant-in-aids from the Ministry of Education, Culture, Sports, Science, and Technology (MEXT) of Japan (20H01944 and 23H01211).

\end{document}